# Study on Extreme Precipitation Trends in Northeast China Based on Non-Stationary Generalized Extreme Value Distribution

Fangxiu Meng[1], Peng Liu [2], Kang Xie[3], Huazhou Chen[1], Yirong Jiang[1]

1College of Science, Guilin University of Technology, Guilin, China

2School of Atmospheric Sciences, Nanjing University of Information Science and Technology, Nanjing, China

3School of Mathematics and Information Sciences, Nanjing Normal University of Special Education, Nanjing, China

**Abstract**

Northeast China is the main food productive base of China. The extreme precipitation (EP) event seriously impacts agricultural production and social life. Given the limited understanding of the EP in Northeast China, we investigate the trend and potential risk of the EP in Northeast China (107 stations) during 1959–2017, especially in early and later summer. For the first time, the non-stationary generalized extreme value (GEV) model is used to analyze the trend and potential risk of the EP in Northeast China. Moreover, the mechanisms of EP trends over Northeast China in early and later summer were studied separately. Negative trends dominate EP in early summer but positive trends prevail in later summer. It is reasonable to discuss separately in the two periods. Meanwhile, all return levels are shown apparently increasing trends in EP in early summer, corresponding to more frequent EP events. Nevertheless, in later summer, the 2-year return level decreases in location parameter diminish slightly, and the rare EP (20-, 50- and 100-year return levels) slightly increase with scale parameter. Also, our results show that normal EP occurs frequently in Liaoning Province, and extreme EP is more likely to occur in Jilin Province and Heilongjiang Province. The increase of EP in early summer is mainly influenced by the northeast cold vortex. However, in later summer, the effect of cold air on EP is stronger in Northeast China, which gives a clear explanation that the EP does not increase. This study analyzed the trends and mechanism of return level and EP, which is beneficial for the development of policy strategies.

# 1. Introduction

According to the Sixth Assessment Report of the Intergovernmental Panel on Climate Change (IPCC), the past 20 years have been the warmest period since the early 20th century.The global warming rate over the last 50 years (since the 1970s) is greater than that of any other 50-year period in the last 2,000 years. With climate warming, extreme heat events (including heat waves) will continue to increase and intensity at the global and continental scales, and extreme cold events will decrease and weaken. In addition, heavy precipitation events are likely to become stronger and more frequent, resulting in increased flood and waterlogging complex events in some coastal and estuarine areas (Fan et al., 2021; IPCC, 2021; Zhou et al., 2021). Such extreme events can cause tremendous losses of life and property and affect agricultural production, the economy and environment. Therefore, studying the trends and occurrence probability of extreme events is instructive for improving the forecasting and early warning capabilities for extreme events and mitigating risks associated with them.

Many studies have shown clear increasing trends in extreme precipitation (EP) in the United States, Australia, China, India, and other regions (Suppiah and Hennessy, 1998; Groisman et al., 2001; Roy and Balling, 2004; Liu et al., 2005; Chen et al., 2014; Hartmann et al., 2014). For instance,Hartmann et al. (2014) found that significantly increasing trends in EP were mainly detected in the very high mountainous regions in the east and in the north of the Indus basin. This is influenced by the western disturbance transporting moisture from the Mediterranean Sea and the Atlantic Ocean. Chen et al. (2014) identified positive trends in the Island of Hawaii. A positive relationship is found between the EP and Southern Oscillation Index (SOI), implying greater extreme events during La Nina years. However, the spatio-temporal variability of the EP in different regions is significant, and the influencing factors are also different. Global precipitation exhibits a widespread and obvious increasing trend. But, the spatial consistency of the EP variability trends is poor (Alexander et al., 2006). The EP over the western part of the Indus River Basin exhibits a decreasing trend. On the contrary,the EP in the east and north of the Indus River basin presents a noticeable increasing trend (Hartmann et al., 2014). The trend of EP is studied by multiple extreme indices. Sheikh et al. (2015) showed that the trends of extreme indices in the Indo-Gangetic Plain are different. Rainy days decrease in the Indo-Gangetic Plain. Similarly, Ren et al. (2017) and Zilli et al. (2017) showed that the trend of rainy days is also decreasing in Southwest China and Myanmar and in large urbanized areas of Sao Paulo state separately. Summer precipitation in Northeast China shows a decreasing trend, with interannual variability periods of 14 years and 2–4 years (Han et al., 2005). Therefore, it is particularly important to investigate the EP trend in different regions.

Moreover, most previous studies on the precipitation in China focused on precipitation anomalies at the national scale or in southern China, and relatively few studies were conducted on the summer EP in Northeast China, especially in early and later summer. As one of the most important food production bases in China, Northeast China is of great importance to national food security. Summer is the critical period for crop growth in Northeast China and also the concentration of the EP in this region. In recent years, EP events have occurred frequently in Northeast China, which has had a serious impact on agricultural production and social life. For example, the highest 1-day precipitation record of more than 135 mm at Jilin Province on July 28, 2010, with the direct economic losses exceeding 6.6 billion RMB. The same year, there were 24 floods in Liaoning Province. Among them, heavy precipitation occurred 6 times (Li, et al.,2012). According to the statistics of the 2012 climate disaster bulletin of Liaoning Province, four regional rainstorms successively occurred in Liaoning Province in 2012, of which the "8.04" flood was the most serious rainstorm. Based on the statistics of the National Disaster Reduction Center, since September 2018, some areas of Northeast China have been hit by heavy precipitation, resulting in 100 thousand mu of affected crops and more than 9 billion RMB of economic losses. All of these instances show that EP events caused widespread damage to life, agriculture, and property. Consequently, a comprehensive analysis of the long-term spatio-temporal trends and the potential risks of the summer EP in Northeast China is of strategic significance for the development of food bases in Northeast China and the guarantee of national food security.

The extreme value model is widely used in meteorological and hydrological to research the statistical characteristics of extreme events (Kharin and Zwiers, 2005; García et al, 2007; Katz et al.,2013; Chen and Chu, 2014; Tan, 2017; Xavier et al., 2020; Zhang et al., 2020; Wu et al., 2021). The generalized extreme value (GEV) model is used to describe the characteristics of maximum precipitation and streamflow (Katz et al.,2013). Considering the introduction of additional climate indices in the non-stationary GEV model, the relationships between climate indices and extreme events were discussed (Chen and Chu, 2014; Tan, 2016; Wu et al., 2021; Zhang et al., 2020). According to the non-stationary GEV model (changing with time), the return levels of extreme events were estimated and it is found that the change of EP is related to the location and scale parameters (Kharin and Zwiers,2005; García et al,2007; Chen and Chu, 2014；Xavier et al., 2020). However, it was rarely used to study the trends of EP and return level.

The study aims to determine the trends and potential risks of the EP in Northeast China in early and later summer during 1959–2017. The EP trends are investigated by the Mann-Kendall (MK) test. As a supplement, a non-stationary GEV model is used in the research on the trends and potential risks of the EP over Northeast China for the first time. Additionally, the effects of non-stationary model parameters on the return levels are also examined. These findings help to understand the trend and occurrence probability of the EP over Northeast China in early and later summer and provide a reference for risk management.

The remainder of this paper is organized as follows. Study area and data are briefly described in Section 2. In Section 3, the principles of the research methods are introduced.

The main results are presented in Section 4. Section 5 discusses the main conclusions and provided the reference for policy strategies and risk management.

**2. Study area and data**

The study area is Northeast China, located in the region of 38°N–53°N, 115°E–135°E (Figure 1), including Liaoning Province, Jilin Province, Heilongjiang Province, and Hulun Buir City, Hinggan League, Tongliao City and Chifeng City of eastern Inner Mongolia (Du et al., 2013; Wu et al.,2021). Northeast China is located in the middle-high latitudes on the east coast of Eurasia, and its climate belongs to the temperate monsoon climate. It is one of the most sensitive regions to global climate change. The spatio-temporal distributions of extreme precipitation (EP) are highly variable.

The daily precipitation records from 116 weather stations in Northeast China during 1951–2017 are obtained from the National Meteorological Information Center of the China Meteorological Administration. The non-stationary GEV model needs complete records gathered over a long time period. In order to preserve as many sites as possible and observe continuously, the following data quality control principles are formulated. If the proportion of missing data from a meteorological station exceeds 5% within a year, the year will be deleted. If the total missing measurement ratio of meteorological station data exceeds 5% of the total records, the station will be deleted. For the percentage of missing observations is less than 5%, we compare the station with nearby stations with complete records and check for abnormally values (Chen et al., 2014). If there is no abnormally value, the missing data shall be supplemented by the historical mean value of the same period. For the abnormally value, we confirm the authenticity of the abnormally value by looking up the data. If it is not true, we replace the abnormal data with the historical mean value of the same period. Finally, the data from 107 stations during 1959–2017 are used in this study to ensure data consistency and completeness.

Previous studies on the EP in Northeast China were mostly conducted throughout the summer (Han et al., 2005; He et al., 2006). However, the intra-seasonal variability of the EP is noticeable during summer (Shen et al., 2011). Therefore, this study is performed in early summer (May–June) and later summer (July–August). According to daily precipitation records statistics in Northeast China, the annual maximum 24-hour accumulated precipitation in Northeast China occurs from May to August. Hence, in early summer, the EP is defined as the maximum 24-hour accumulated precipitation from May to June each year. We select the maximum 24-hour accumulated precipitation from July to August each year as the EP during the later summer (Gao et al., 2016; Jeon et al., 2016; Tan, 2016; Zhang et al., 2020).

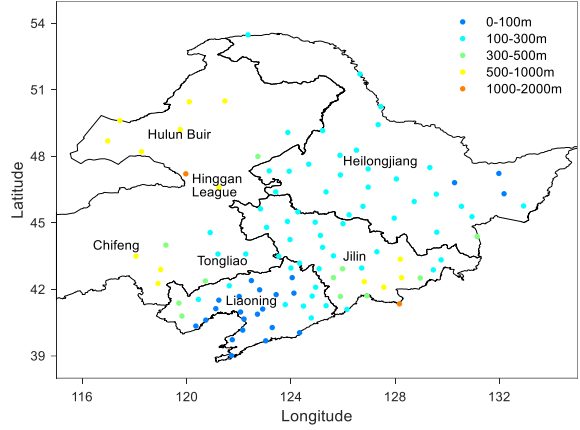

**Figure1.** Spatial distribution of study area of EP over Northeast China

## 3. Methods

The EP variability is investigated by using the MK test. Additionally, a non-stationary GEV model is adopted for the first time to analyze the trends and potential risks of the EP in Northeast China.

### 3.1 The Mann-Kendall test

The MK test was first proposed by Mann (1945) and implemented by Kendall (1975), which is widely used in trend testing and mutation detection. It does not require a fixed data distribution and is not disturbed by a few outliers. In addition, the MK method does not require high data quality. This method is extensively adopted for trend analysis in hydrology and meteorology (Wei, 2007). The specific steps are shown as follows.

For the time series $X = \{x_1, x_2, x_3 \cdots x_n\}$, an ordered series can be constructed as Equation (1).

$$S_k = \sum_{i=1}^{k} r_i \quad (k = 2, 3, \ldots, n) \qquad (1),$$

where $r_i$ can be calculated by Equation (2).

$$r_i = \begin{cases} +1, & \text{if } x_i > x_j \\ 0, & \text{if } x_i \leq x_j \end{cases} \quad (j = 1, 2, \ldots, i) \qquad (2),$$

where $r_i$ denotes the number of times when the value at time $i$ is greater than that at time $j$. $S_k$ indicates the cumulative value of $r_i$.

Assuming that the time series $S_k$ is randomly independent, the statistic can be defined as Equation (3).

$$\text{UF}_k = \frac{S_k - E(S_k)}{\sqrt{\text{Var}(S_k)}} \quad (k = 1, 2, \ldots, n) \qquad (3),$$

where $UF_1 = 0$, $E(S_k)$ and $Var(S_k)$ represent the mean and variance of $S_k$, respectively. Assuming that $x_1, x_2, x_3 \cdots x_n$ are independent of each other and have the same continuous distribution, $E(S_k)$ and $Var(S_k)$ can be calculated by Equation (4) and (5).

$$E(S_k) = \frac{n(n-1)}{4} \quad (4),$$

$$Var(S_k) = \frac{n(n-1)(2n+5)}{72} \quad (5).$$

The significance of the trend is tested by using $UF_k$ value. Generally, positive $UF_k$ value represent an upward trend, and negative ones mean a upward trend. The trend is considered significant if the absolute of $UF_k$ is greater than 1.96. Repeat the above procedure for the time series in reverse order, i.e., $\{x_n, x_{n-1}, x_{n-2}, \cdots, x_1\}$, and calculate the statistic $UB_k$ to make the $UB_k$ equal to $-UF_k$. Based on the intersection points of the UF and UB curves, mutation years are detected.

### 3.2 Non-stationary generalized extreme value model

According to the definition of extreme events, if the EP represented by $\{y_1, \ldots, y_N\}$ is independently and identically distributed with a common cumulative distribution function, $M_n = \max\{y_1, \ldots, y_N\}$ obeys one of three distributions: Gumbel, Fréchet and Weibull distributions. We refer to these three distributions collectively as the GEV distribution (Coles, 2001), as shown in Equation (6).

$$F(m) = \exp\left\{-\left[1 + \xi\left(\frac{m-\mu}{\sigma}\right)\right]^{-1/\xi}\right\}, 1 + \xi\left(\frac{m-\mu}{\sigma}\right) > 0 \quad (6),$$

where $\mu$, $\sigma$ ($> 0$) and $\xi$ indicate the location, scale and shape parameters, respectively. Positive, zero, and negative values of the shape parameter represent different distributions.

To analyze the variation trends of the EP, we allow the GEV parameters (location and scale) to vary with time in this study (Katz, 2013). The five common types of EP trends are shown in Table 1.

**Table 1.** Trend types of non-stationary GEV model

| Trend types | Parameter forms |
|---|---|
| $\mu$ linear trend, $\sigma$ no trend (AL) | $\mu(t) = \mu_0 + \mu_1 t, \sigma(t) = \sigma_0$ |
| $\mu$ parabolic trend, $\sigma$ no trend (AP) | $\mu(t) = \mu_0 + \mu_1 t + \mu_2 t^2, \sigma(t) = \sigma_0$ |
| $\mu$ no trend, $\sigma$ linear trend (BL) | $\mu(t) = \mu_0, \sigma(t) = \sigma_0 + \sigma_1 t$ |
| $\mu$ no trend, $\sigma$ parabolic trend (BP) | $\mu(t) = \mu_0, \sigma(t) = \sigma_0 + \sigma_1 t + \sigma_2 t^2$ |
| $\mu$ linear trend, $\sigma$ linear trend (DL) | $\mu(t) = \mu_0 + \mu_1 t, \sigma(t) = \sigma_0 + \sigma_1 t$ |

## 3.3 *T*-year return level

The *T*-year return level is defined as the reciprocal of *P*, i.e., $P = \frac{1}{T}$, which is the probability of exceeding the annual extremes in any given year. It is often referred to as a *T*-year event in hydrology (Filliben, 1975; Coles, 2001). The relationship between the return period ($T$) and the associated return level ($x_T$) can be expressed as Equation (7).

$$F(x_T) = P(M_y \leq x_T) = 1 - \frac{1}{T} \qquad (7),$$

where $x_T$ denotes the $(1 - \frac{1}{T})$ quantile of EP probability distribution for any year. $M_y$ represents the EP.

According to the principle of the minimum AIC value, the best-fit models for the EP in Northeast China are selected under non-stationary conditions, and the time-varying trend types of the location and scale parameters are determined. Then, the time-varying *T*-year return level $x_T$ is obtained by Equation(8).

$$x_T = \mu - \frac{\sigma}{\xi}\left[1 - \left\{-\log\left(1 - \frac{1}{T}\right)\right\}^{-\xi}\right] \qquad (8).$$

## 4. Results

### 4.1 Trends of extreme precipitation using the Mann-Kendall test

Through the MK method, the significance test for the EP trends in Northeast China is performed at the 0.05 significance level (Figure 2). According to the UF curve, in early summer, the EP shows a significant increasing trend since the 1970s. Especially after the 1980s, the EP trend is more evident, which passed the test at 99% confidence. On the contrary, since the 1960s, a significant decreasing trend dominates the EP in later summer. Especially from the 1970s to the 1990s, the EP trend is more significant at 0.01 level. Based on the intersection points of the UF and UB curves, it is determined that the increase of the EP in early summer shows a mutation, specifically from 1972 to 1995. The increasing trends of EP were more evident after the mutation in 1972 and slightly decreased after the mutation in 1995.The decrease of the EP in later summer also presents a mutation, specifically from 1966 to 1998. The slopes of EP reduction increased after a sudden change in 1966, and became smooth after 1998. These results are highly similar to previous studies (Lin et al., 2021).

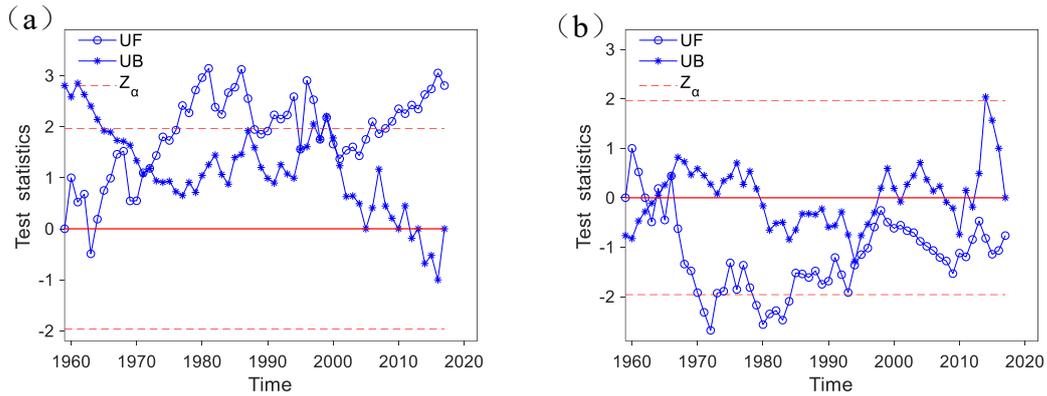

**Figure 2.** MK test of EP in NEC (a) early summer and (b) later summer

The spatial distributions of the EP trends in Northeast China in early and later summer are shown in Figure 3. The results show that the EP at 90 stations shows an increasing trend over time. The number of stations with an increasing EP trend approximately 84.11% of all stations. In particular, there are 13 stations with increasing EP trends that are significant at the 5% level and even passed the test at 99% confidence. These stations are mainly located in Liaoning, Jilin, and Heilongjiang Provinces. Among them, the stations with a significant increase in EP are mostly located in Heilongjiang Provinces, especially in Songhuajiang-Nenjiang River Basin, where precipitation is abundant and is the maximum EP center. In later summer, decreasing trends prevail the EP at 53 stations. The number of stations with decreasing EP trends and increasing EP trends is similar. However, the upward EP trends did not pass the significance test. The decreasing EP trends in Tongliao City and the northwest of Jilin Province are significant at the 5% level, and these regions are located in the Liaohe River and Songhua River basins.

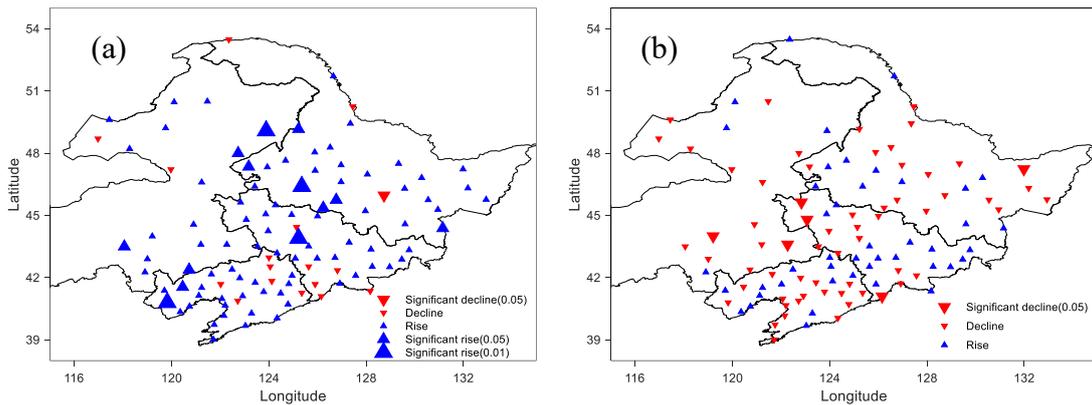

**Figure 3.** Spatial distribution for the EP trends by using the MK test in early summer (a) and later summer (b)

## 4.2 Extreme precipitation Trend obtained by the non-stationary GEV model

### 4.2.1 Trends of non-stationary GEV parameters

To complement the results obtained from the MK test, we fit time-dependent GEV model to analyze the EP trends in Northeast China. Spatial distributions of the trends of the location and scale parameter are presented in Figure 4. Among them, the location parameter $u_1$ corresponds to the mean value of EP, while the scale parameter $\sigma_1$ represents the variance of EP. In early summer, positive trends dominate two parameters, corresponding to the enhancement of EP. The trends are consistent with the results obtained by the MK test (Figure 2). Based on the results from the likelihood ratio test, location parameter $u_1$ at about 82.24% of the stations is positive. Among them, the $u_1$ trend is significantly positive at the 0.05 level at 22.73% of the stations, and these stations are mainly located in the southwest of Liaoning Province, Tongliao City, the northwest of Jilin Province, and the south of Heilongjiang Province. The trends of scale parameter $\sigma_1$ and location parameter $u_1$ are not always the same. The scale parameter $\sigma_1$ at 92 stations displays a positive trend. In particular, there is an obvious increase in 35 stations, and these stations are also mainly located in the southwest of Liaoning Province, Tongliao City, the northwest of Jilin Province, and the south of Heilongjiang Province. During later summer, the trend of location parameter $u_1$ is not obvious and slightly decrease, corresponding to the decline of EP. The opposite trend behavior in scale parameter $\sigma_1$ shows a positive trend. A significantly negative trend prevails location parameter $u_1$ at the 0.05 level in Tongliao City, northwestern Jilin Province, and central Heilongjiang Province. Scale parameter $\sigma_1$ at 87 stations presents a positive trend, especially at 37 stations, with a significant trend. Note that these stations are also mainly located in Tongliao City, northwestern Jilin Province, and central Heilongjiang Province. These results are highly similar to the EP trends obtained by the MK test (Figure 3).

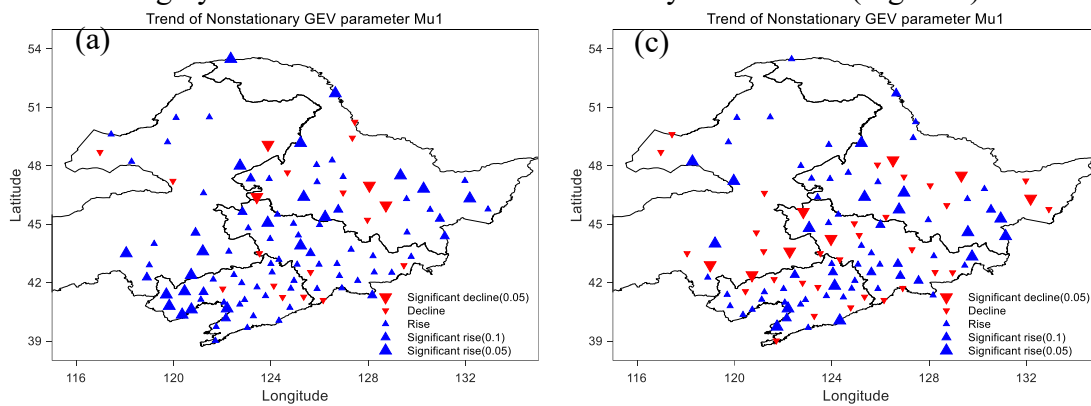

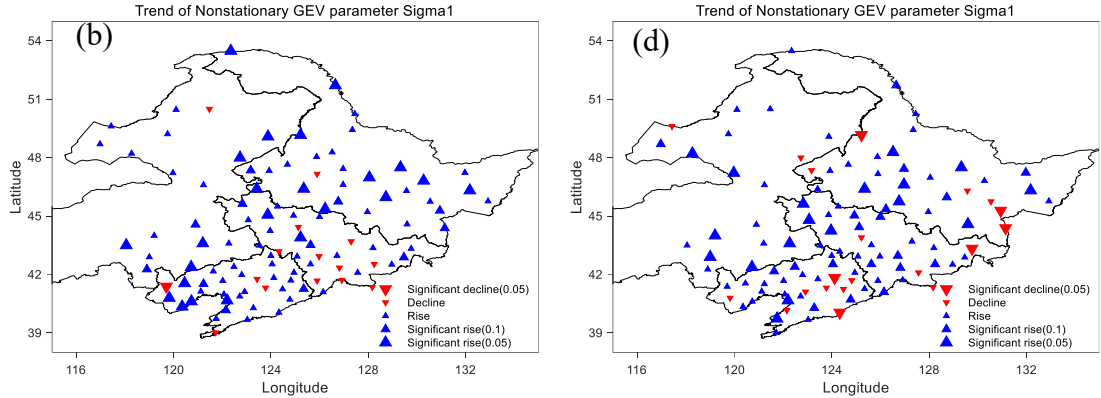

**Figure 4.** Spatial distribution of trend of location parameter $u_1$ and scale parameter $\sigma_1$ for EP at NEC in early summer (a, b) and later summer (c, d)

### 4.2.2 Extreme precipitation trends at the return levels

Figure 5 shows the time series of 2-, 20-, 50- and 100-year return levels of the EP in Northeast China in early and later summer, which are obtained by transforming the standardized data back to the original space. Results show that the EP observations in Northeast China are not constant. Specifically, the clear increasing trends of return levels (20-, 50- and 100-year) are observed in both periods and the 100-year return level increased at the highest rate. Particularly in early summer, it increases from 84.05 mm in 1969 to 95.31 mm in 2017, with a linear variability of 1.9 mm (10 years)$^{-1}$. EP events are more frequent, especially in early summer. For instance, the precipitation threshold value with 100-year in 1959 that has occurred on once every 51 to 52 years by 2017. Similarly, the rare EP (50-year return level) in 1959 that has become relatively comment (22-23 years) in 2017. The 2-year return levels are relatively stable in both periods, showing slight increase in early summer with a linear variability of 0.69 mm (10 years)$^{-1}$. Nevertheless, in later summer, there is a slightly decreasing trend in the 2-year return level. Specifically, the EP in the 2-year return level increased from 25.96 mm in 1959 to 29.96 mm in 2017 in early summer. While in later summer, it decreases from 54.17 mm in 1959 to 52.38 in 2017. According to Figure 4, when both location parameter $u_1$ and scale parameter $\sigma_1$ have positive trends, the return level increases with the growth of T. Therefore, in early summer, the growth rates of the 20-, 50- and 100-year return levels are greater than those of the 2-year return level. However, when location parameter $u_1$ shows a negative trend and scale parameter $\sigma_1$ displays a positive trend, the trends of return levels are not the same. There is a slightly decreasing trend in the 2-year return level, while it in the 20-, 50- and 100-year return levels slightly increase. This phenomenon is caused by the different trends of the two parameters. The location parameter $u_1$ has a major effect on the 2-year return level. While, the scale parameter $\sigma_1$ contribute more to the 20-, 50- and 100-year return levels.

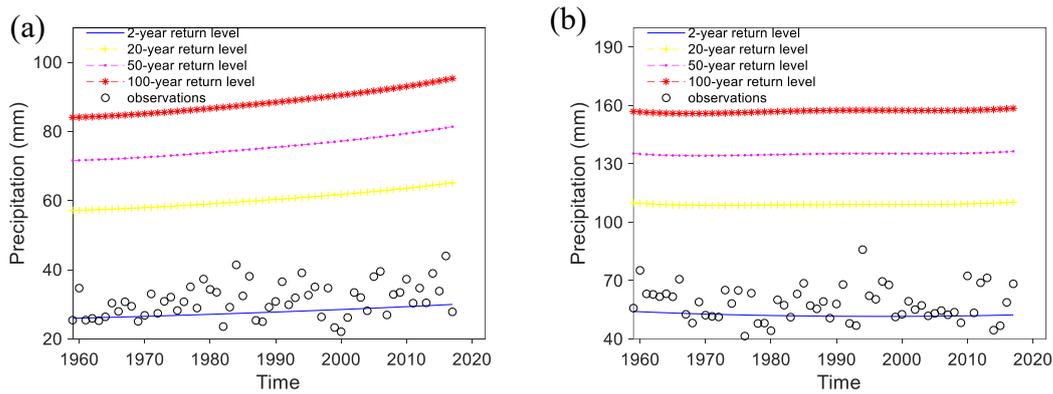

**Figure 5.** Time series of 2-, 20-, 50- and 100-year return levels in early summer (a) and later summer (b)

Furthermore, the trends of the normal EP (2-year return level) and extreme EP (100-year return level) are tested by the MK test and aim to study the spatial variability of normal and extreme EP in Northeast China, as shown in Figure 6. The results suggest that in early summer, the normal and extreme EP both show significantly increasing trends, and the spatial distributions of these EP trends are consistent. However, in later summer, there is no remarkable difference in the number of stations between decreasing EP trends and increasing EP trends. During early summer, 76.64% of stations show an increasing trend in normal EP, among them 90.24% are significantly at 99% confidence. The difference is very small between the number of stations with decreasing EP trends and increasing EP trends. Approximately 50.47% of all stations show decreasing trends, among them 85.19% are significantly at the 1% level. The extreme EP increases significantly at 53 stations, which are located in southwestern Liaoning Province, central Jilin Province, western Heilongjiang Province and Hulun Buir, i.e., the Liaohe, Mudanjiang and Nengjiang River basins, where precipitation is abundant and prone to EP events. The spatial distribution of the normal EP trends is consistent with that extreme EP. In general, the normal and extreme EP over Northeast China show significantly increasing trends in early summer, and most of Northeast China faces an increasing trend of EP risk, indicating that EP events would become more frequent (Figure 5). However, in later summer, this trend is not obvious. It is necessary to focus on the stations in the Liaohe, Mudanjiang and Nenjiang River basins and strengthen the precautions against EP impacts.

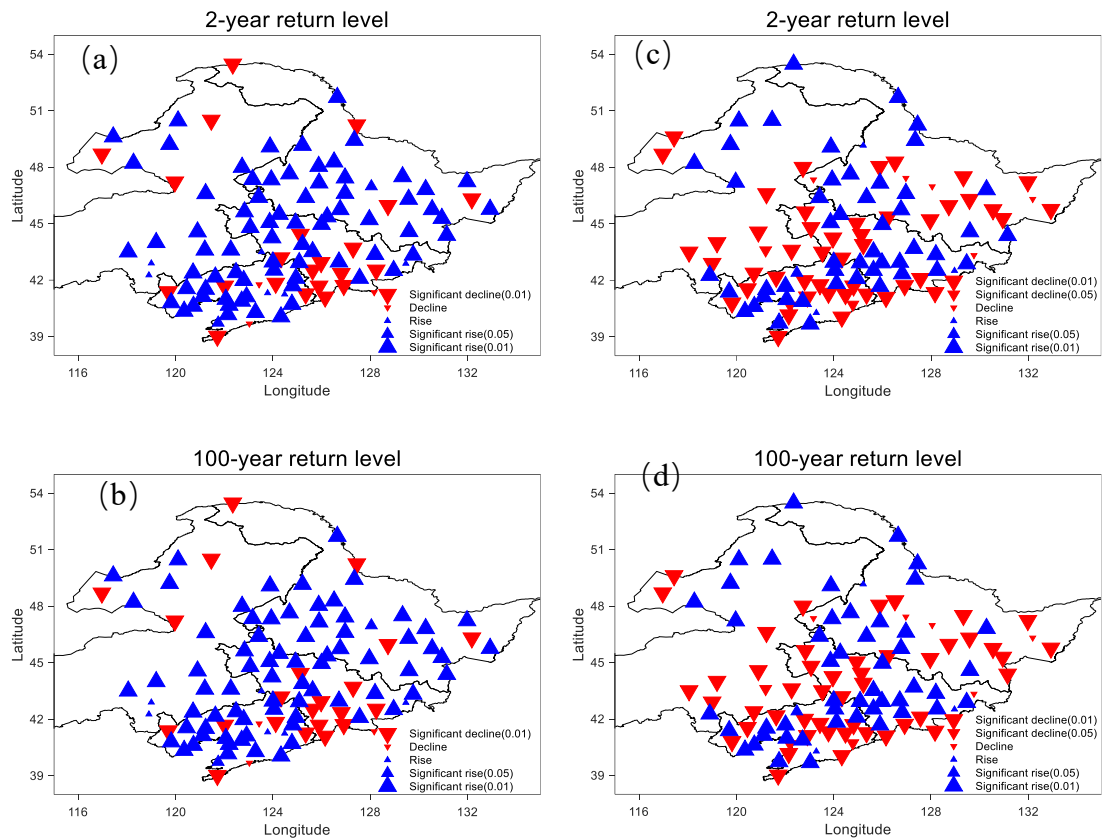

**Figure6.** Spatial distribution of MK test for normal EP (a, c) and extreme EP (b, d) in early summer (a, b) and later summer (c, d)

## 4.3 Potential risk of the extreme precipitation in Northeast China

Using the non-stationary GEV model, we assess the potential risk of the EP based on the normal EP and extreme EP with rainstorms, heavy rainstorms, and above the regional average (Figure 7). According to the national standard of rainfall grade, it is considered that rainstorm occurs when the accumulated precipitation reaches 50-99mm in a day. And a heavy rainstorm occurs when the accumulated precipitation reaches 100 to 200mm in a day. Based on the non-stationary GEV model, the EP corresponds to the 59-year time series. And the probability of rainstorms in EP is the frequency of 50-99 mm in the time series. Other probabilities are similar to calculations. In early summer, the probability of rainstorms in normal EP is zero in Northeast China, which does not suggest that the occurrence of rainstorms is impossible. In later summer, rainstorms can occur in most of Northeast China, especially in Liaoning Province, which has the possibility of rainstorms of greater than 0.7 and is mainly influenced by the Liaohe and Yalujiang River basins. During early summer, the rainstorm probability in extreme EP is 0.8 and above in most of Northeast China. Especially in Tongliao City, Chifeng City, and eastern part of Northeast China, the probability of heavy rainstorms is 0.7 and above. The southeast and west of Northeast China have a high probability of exceeding the regional average EP. In later summer, the probability of rainstorms in the north and northwest parts reaches 0.7 and above, and the probability of heavy rainstorms is 0.7 and above in most of Northeast China. In Liaoning Province, Tongliao City, and Chifeng City, the probability of rainstorms reaches 0.7 and above, exceeding the regional average EP. In general, the extreme EP is dominated by rainstorms in early summer and heavy rainstorms in later summer. Note that normal EP occurs frequently in Liaoning Province, and extreme EP is more likely to occur in Jilin Province and Heilongjiang Province. The EP values in Liaoning Province, Tongliao City, and Chifeng City are more than that in the whole region.

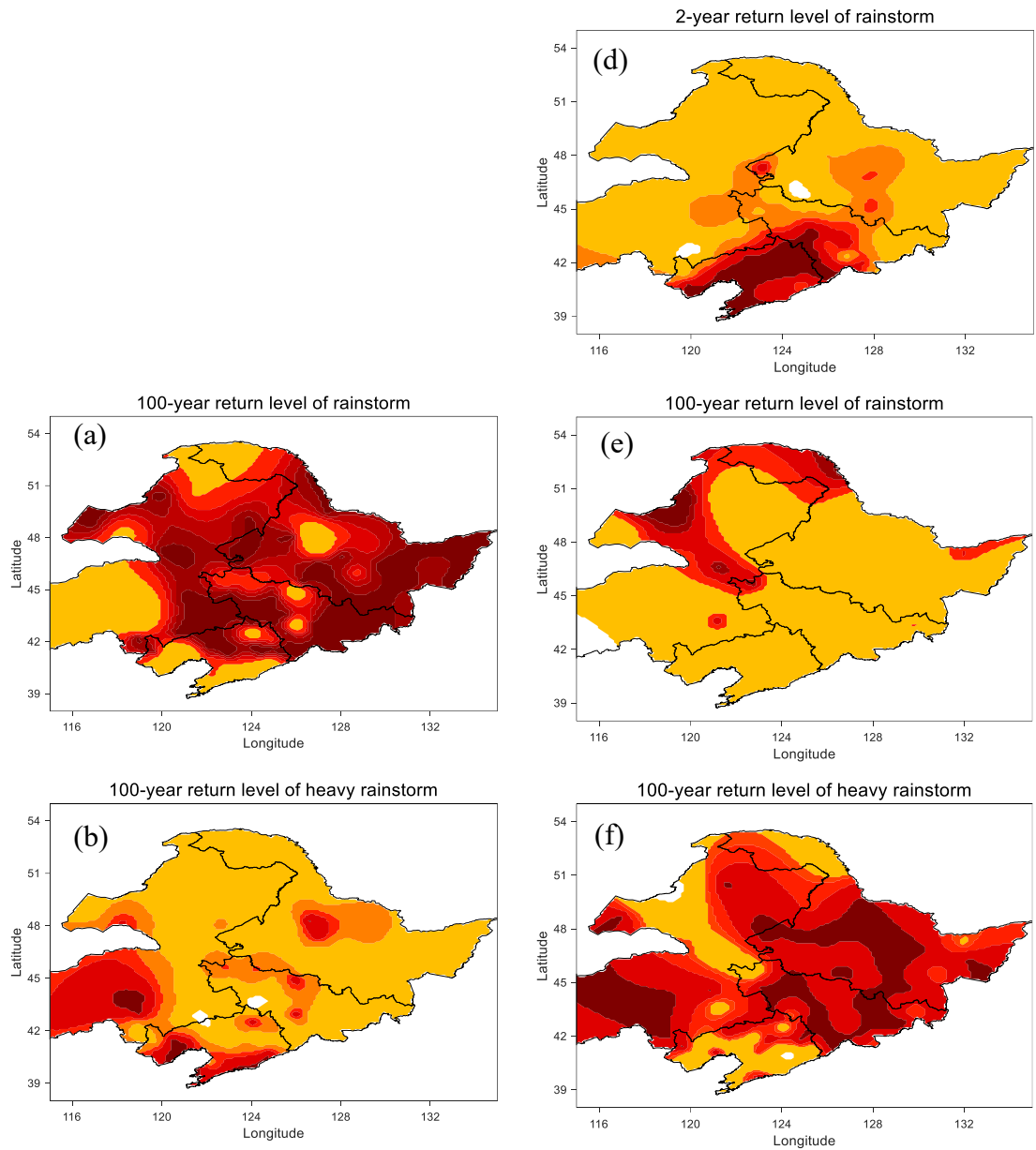

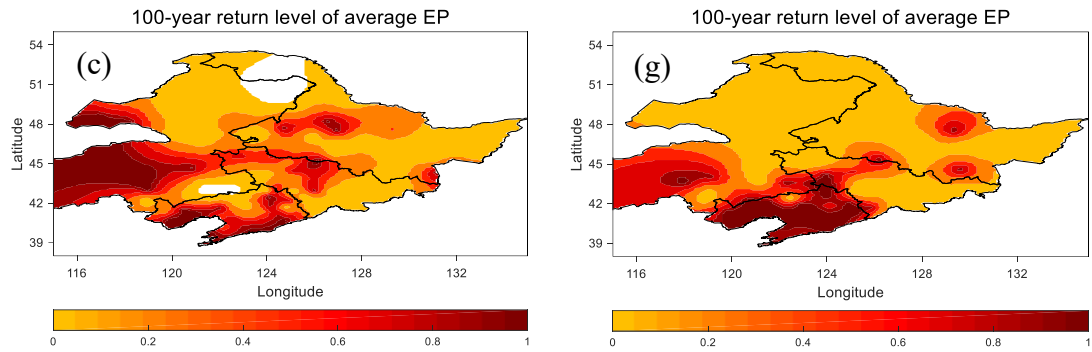

**Figure 7.** Probability spatial distribution of normal and extreme EP of rainstorm (a, e), heavy rainstorm (b, f) and higher than regional average EP (c, g) in early (left) and later summer (right)

## 4.4 The mechanism of the extreme precipitation in Northeast China

Based on the conclusion in section 4.1, the trend of EP changes in early and later summer in Northeast China underwent two turns from 1959 to 2017. Based on Kriging interpolation method, we analyze the relationships between EP and large-scale circulation in early and later summer in Northeast China to study the impact mechanism of extreme precipitation trends. According to the mutation years of early and later summer (Figure 2), the early summer period of EP is divided into two periods: 1972-1995 (early summer period 1) and 1996-2017 (early summer period 2). Similarly, the later summer period of EP can also be divided into two stages: 1966-1998 (later summer period 1) and 1999-2017 (later summer period 2). Figure 8 and 9 show the correlation coefficients between EP and 500 hPa height field, 850 hPa specific humidity, and wind field in early and later summer respectively. In addition, the MK tests of the circulation field are shown in Figure 10.

As shown in Figures 8 and 9, the EP is significantly positively correlated with the height and the specific humidity field in both periods 1. While in period 2, it is clearly negatively correlated with the height field. In early summer period 1, the EP is mainly affected by the easterly current of the anticyclone circulation, which is conducive to the transport of warm-moist air from the eastern sea of Northeast China to the surface, with the amount of water vapor. During early summer period 2, under the condition of the easterly current of the cyclone circulation and prevailing southerly wind, a large amount of warm-moist air in the Bohai Sea and Yellow Sea moved northward, leading to EP events. There is an obvious cyclonic circulation over the northeast, corresponding to the northeast cold vortex on 500hPa. In later summer, the EP is mainly controlled by the south of the temperate anticyclone, and the influence is smaller in period 2. According to Figure 10, Significant increasing trends of the height field were detected in both periods. During later summer, the range of the height field in the northeast is larger and the cold air and wind in the north are stronger than that in the early summer. That means, in later summer, cold air dominated the EP in Northeast China, which inhibited the effect of water vapor.

Therefore, there is no increasing trend of EP in later summer. On the contrary, in early summer, cold and warm-moist air is active. The warm and moist air moves northward while the cold air moves southward. Thus, the stronger moisture convergence over Northeast China, gives rise to increasing in EP in early summer.

Overall, in the two periods, the reason for the different trends of EP is the different characteristics of the circulation field. The northeast cold vortex is the main influencing factor of EP in early summer, especially in period 2. The stronger moisture convergence over Northeast China, is conducive to the increase in EP in early summer. Nevertheless, the south of the temperate anticyclone controls EP in later summer. The cold air dominated the EP in Northeast China, which inhibited the effect of water vapor. So the increasing trend of EP is not obvious in later summer. These results are highly similar to previous studies (Shen et al., 2011; Lin et al., 2021; Wu et al.,2021).

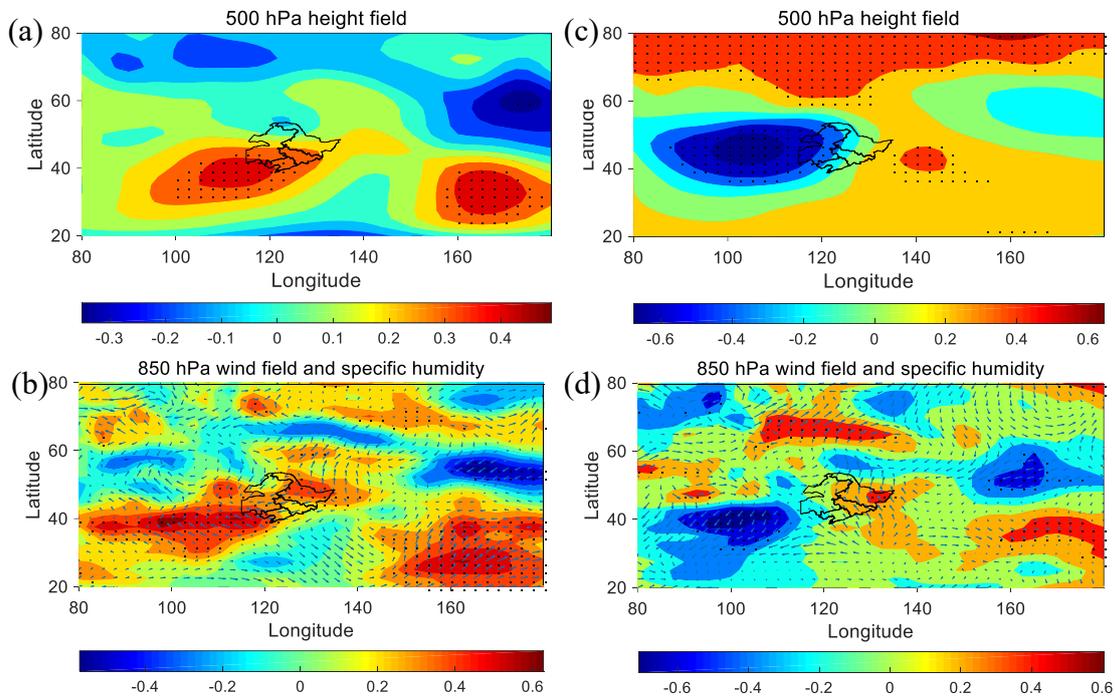

**Figure 8.** Correlation coefficients between EP and 500 hPa height field (a, c), 850 hPa specific humidity field and wind field (b, d) in early summer period 1 (left) and early summer period 2 (right). Dotting means correlation coefficients significance at the 5% level.

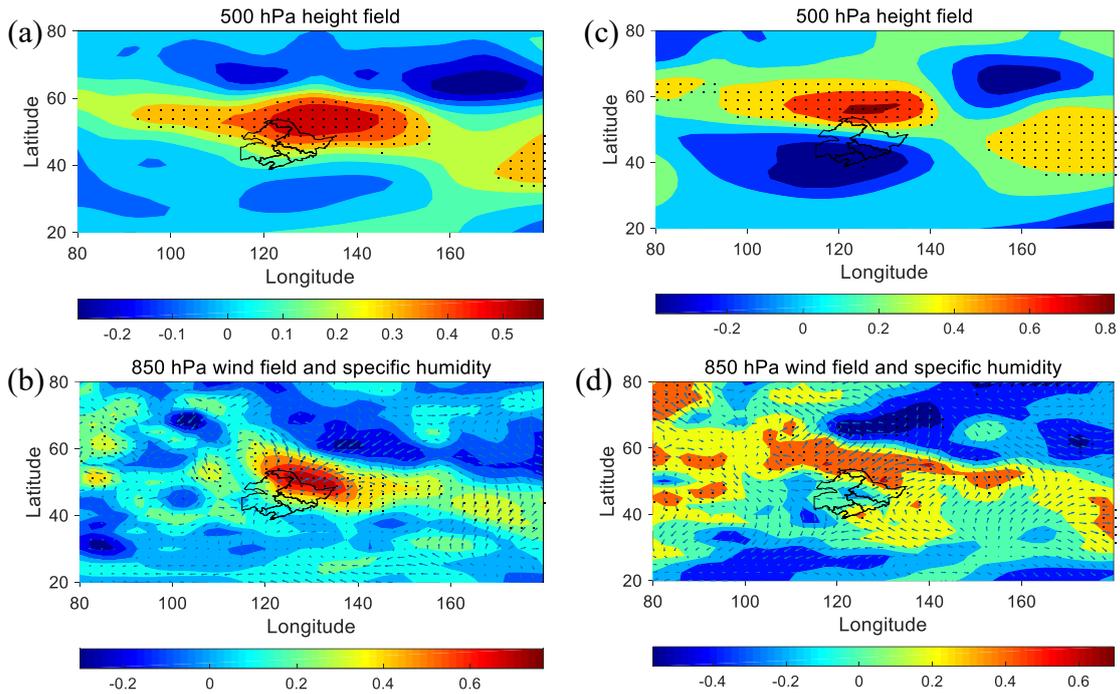

**Figure 9.** Correlation coefficients between EP and 500 hPa height field (a, c), 850 hPa specific humidity field and wind field (b, d) in later summer period 1 (left) and later summer period 2 (right). Dotting means correlation coefficients significance at the 5% level.

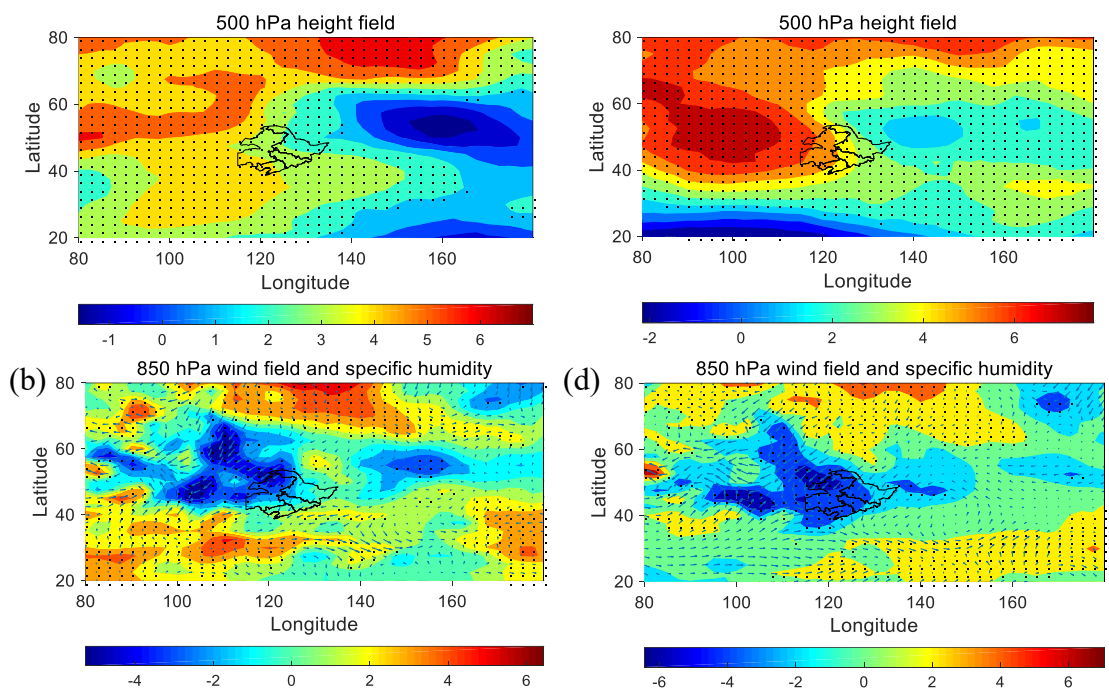

**Figure10.** Spatial distribution of MK test for 500 hPa height field (a, c), 850 hPa specific humidity field and wind field (b, d) in early summer (left) and later summer (right). Dotting means correlation coefficients significance at the 5% level.

## 5. Conclusions

In this study, the trends and potential risks of EP over Northeast China are analyzed. Firstly, the MK test is applied to detect mutations and trends of EP. As a supplement, a non-stationary GEV model is used to investigate the trend and potential risk of the EP over Northeast China for the first time. Meanwhile, the relationships of non-stationary model parameters with return levels are discussed. Moreover, the mechanism of EP in Northeast China also is studied.

Based on results obtained from the MK test, negative trends dominate EP in early summer but positive trends prevail in later summer. In addition, there are two mutation years in early summer (1972, 1995) and later summer (1966, 1998) respectively. This conclusion is consistent with previous studies (Lin, et al.,2021). Therefore, early and later summer is divided into two periods to discuss the mechanism of EP trends. Then using a non-stationary GEV model, all return levels are shown apparent increasing trends in EP in early summer, and the growth rates of the 2-year return level is less than that those of the 20-, 50- and 100-year return levels. The results indicated that EP events are more frequent. For instance, the extreme EP (100-year return level) in 1959 that has become relatively comment (51-52 years) in 2017. In later summer, 20-, 50- and 100-year return levels also show slightly increase trends. Conversely, there is a slightly decreasing trend in the 2-year return level. This phenomenon is caused by the different trends of location parameter $u_1$ and scale parameter $\sigma_1$. The location parameter $u_1$ has a major effect on the 2-year return level. While, the scale parameter $\sigma_1$ contributes more to the 20-, 50- and 100-year return levels. Additionally, the potential risks of normal and extreme EP are assessed. Note that in early summer, the rainstorm dominates extreme EP in most of Northeast China, especially in Tongliao City, Chifeng City, and southeastern Northeast China, which are the key areas for the Northeast cold vortex. And heavy rainstorm prevails in later summer, especially in Jilin Province and Heilongjiang Province. That means normal EP occurs frequently in Liaoning Province, and extreme EP is more likely to occur in Jilin Province and Heilongjiang Province. Besides, the EP values in Liaoning Province, Tongliao City, and Chifeng City are more than that in the whole region. Those results provide a reference for engineering design.

To understand the physical mechanism of EP in Northeast China, the correlation coefficients between EP and 500 hPa height field, 850 hPa specific humidity, and wind field are calculated. The trends of the circulation filed are detected by the MK tests at the same time. The effect of the Northeast cold vortex on EP is significant in early summer. The stronger moisture convergence in Northeast China, corresponds to the increase of EP in early summer. However, the south of the temperate anticyclone is the main influencing system of EP in later summer. The cold air dominated the EP in Northeast China, which inhibited the effect of water vapor, corresponding to the increasing trend of EP is not obvious in later summer.

In a word, this study investigates the trends of return level and EP over Northeast

China through a non-stationary GEV model for the first time. Meanwhile, the effect of non-stationary model parameters on return levels is discussed. Furthermore, there is a clear mechanism corresponding to the EP trends in Northeast China. These findings provide references for policy formulation and risk management.

**Acknowledgments**

This research was supported by National Natural Science Foundation of China (1216102 8), and Natural Science Foundation of Guangxi Province (2018GXNSFAA050045). The authors appreciate the associate editor and reviewer for their constructive comments that contributed to improving the manuscript.

**Conflict of interests**

The authors declare no potential conflict of interests.